\documentclass[journal]{IEEEtran}
\usepackage{graphicx}
\usepackage{amsmath}
\usepackage{amssymb}
\usepackage{amsbsy}
\usepackage{multirow}

\begin{document}
%
\title{Maximum Voiced Frequency Estimation: Exploiting Amplitude and Phase Spectra}
%

\author{Thomas Drugman, \textit{Member, IEEE}, Yannis Stylianou, \textit{Senior Member, IEEE}}      

\markboth{IEEE Signal Processing Letters}%
{Shell \MakeLowercase{\textit{et al.}}: Bare Demo of IEEEtran.cls for Journals}

\maketitle

\begin{abstract}
Maximum Voiced Frequency (MVF) is used in various speech models as the spectral boundary separating periodic and aperiodic components during the production of voiced sounds. Recent studies have shown that its proper estimation and modeling enhance the quality of statistical parametric speech synthesizers. Contrastingly, these same methods of MVF estimation have been reported to degrade the performance of singing voice synthesizers. This paper proposes a new approach for MVF estimation which exploits both amplitude and phase spectra. It is shown that phase conveys relevant information about the harmonicity of the voice signal, and that it can be jointly used with features derived from the amplitude spectrum. This information is further integrated into a maximum likelihood criterion which provides a decision about the MVF estimate. The proposed technique is compared to two state-of-the-art methods, and shows a superior performance in both objective and subjective evaluations. Perceptual tests indicate a drastic improvement in high-pitched voices.
\end{abstract}


%


\let\thefootnote\relax\footnotetext{
\\Authors are with Toshiba Cambridge Research Laboratory. \textit{Address: 208 Cambridge Science Park, Milton Road, Cambridge, CB4 0GZ, UK},\emph{Phone: +44 1223 436900}, \emph{Email:} yannis.stylianou@crl.toshiba.co.uk.}

\section{Introduction}
\label{sec:Intro}

During the production of voiced sounds, a variety of models assume the speech signal to contain both periodic and aperiodic components. Two main strategies have been proposed in the literature to control the spectral weighting of these two components. The first one relies on a multiband approach where, for each spectral band, the energy ratio between the periodic and aperiodic contributions is controlled by \emph{aperiodicity measurements}. These measurements can be computed in various ways. In \cite{Yoshimura}, they consist of correlation coefficients calculated in each band, while in \cite{Kawahara01,Raitio_ICASSP11} they are determined based on the ratio between the upper and lower smoothed spectral envelopes. The second approach for spectral weighting assumes the spectrum to be split into two bands: periodic components hold only in the low frequencies, while the aperiodic contributions take place in the higher part of the spectrum. The boundary between these two spectral bands is called the \emph{maximum voiced frequency} (MVF), and it has been used in various speech models such as the so-called multiband excitation vocoder \cite{Griffin88}, the Harmonic plus Noise Model (HNM, \cite{HNM,Erro}) and its variants \cite{qHNM}, the Deterministic plus Stochastic Model (DSM, \cite{DSM}) of the excitation signal.

In the early versions of these vocoders, a fixed constant value of the MVF (generally around 4 kHz) was used for a particular speaker with a given voice quality \cite{IS_DSM, HNM, Erro2}. Recent studies however have reported a gain by considering an explicit dynamic modeling of the MVF in Hidden Markov Model (HMM) based speech synthesis \cite{NoiseModeling,Erro}. In \cite{NoiseModeling}, the use of a dynamic MVF was even found to be slightly preferred over the multiband approach. While current methods of MVF estimation seem to be efficient in speech, some issues were reported in \cite{Singing} for the synthesis of singing voices. For high-pitched voices, MVF was observed to be underestimated which led to an excessive amount of noise after synthesis.

All of these studies motivate the need for a technique which estimates accurately the MVF contour both in speech and singing voices. Very few methods have been proposed for this purpose. In \cite{HNM}, a peak-to-valley (P2V) measure is calculated for all possible harmonic candidates. In \cite{Erro}, the authors make use of the Sinusoidal Likeness Measure (SLM), which can be seen as a localized cross-correlation between the harmonic peak and the spectrum of a pure sinusoid. Based on these measurements for each harmonic candidate, the MVF decision can be taken using various strategies. In \cite{HNM}, a binary decision (into harmonic/non-harmonic) is taken for each spectral peak by applying a threshold to the P2V values, and the MVF is defined as the highest frequency for which the harmonicity criterion is met. In \cite{Erro}, SLMs are scaled between 0 and 1 by a nonlinear function and the MVF is defined by minimizing the Euclidian distance to the ideal SLM profile (i.e. 1s below the MVF and 0s beyond it). Since the resulting contours generally exhibit rapid irrelevant variations, a further post-processing is applied: median filtering in \cite{HNM} and dynamic programming search in \cite{Erro}. In \cite{Ciobanu}, the authors proposed to replace the block of MVF determination based on the binary decisions in the P2V framework, by proposing various other strategies. They however reported a degradation of the perceptual quality after synthesis. Finally, it was proposed in \cite{Zivanovic} to classify spectral peaks as sine or noise based on 3 features describing their bandwidth, duration and frequency coherence.

The goal of this paper is to propose new techniques of MVF estimation. With regard to the state of the art, the contributions of this paper are the following: \emph{i)} we investigate if measurements made directly on phase spectra contain relevant information for MVF estimation, \emph{ii)} we envisage various joint uses of amplitude and phase-based measurements, \emph{iii)} we propose a Maximum Likelihood (ML) criterion as a strategy to derive the MVF decisions, \emph{iv)} contrary to current techniques which make use of some \emph{arbitrary} settings, the proposed method is based on empirical settings derived from an objective study carried out on semi-synthetic signals.


The paper is organized as follows. The proposed approach is detailed in Section \ref{sec:Proposed}. Both objective (Section \ref{sec:Objective}) and subjective (Section \ref{sec:Subjective}) evaluations compare the performance of the proposed technique with those of P2V and SLM methods. Section \ref{sec:Conclu} finally concludes the paper.

\section{The Proposed Technique}
\label{sec:Proposed}

\begin{figure}[!htpb]
  \centering
  \includegraphics[width=0.48\textwidth]{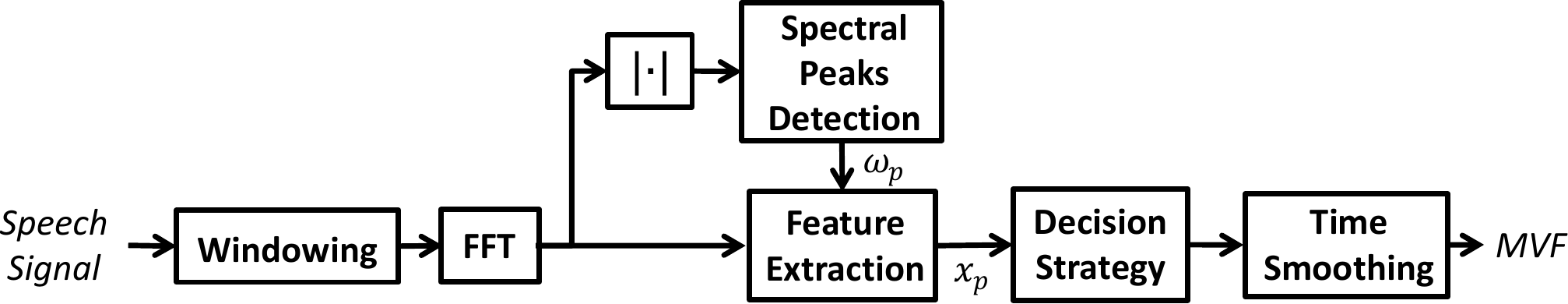}	
  \caption{Workflow of the proposed method.}
  \label{fig:Workflow}  
	\vspace{-16pt}  
\end{figure}

The workflow of the proposed technique is presented in Figure \ref{fig:Workflow}. The steps involved in this workflow are further detailed in the following sections.

\subsection{Windowing}\label{ssec:windowing}
Windowing is essential as it determines the harmonicity properties of the resulting spectra. In all cases, the window length should be proportionnal to the pitch period. In this work, we have used a 4 period-long Hanning window as we found it to be suited for the amplitude spectra to exhibit a good peak-to-valley structure.

\subsection{Spectral Peak Detection}\label{ssec:Peaks}
Starting from an initial estimate of the fundamental frequency $F_0$, the goal of this block is to find the angular frequencies $\omega_p$ of the harmonic candidates up to the Nyquist frequency. A standard peak picking is here applied, where $\omega_p$ is defined as the maximum value in the amplitude spectrum in the neighborhood (10 Hz on each side) of $p\cdot\omega_0$. Then $\omega_0$ is updated to $\frac{\omega_p}{p}$ for the following harmonics. 

\subsection{Feature Extraction}\label{ssec:Features}
For each harmonic candidate $\omega_p$, we propose to extract three types of measurements $x_p$: Amplitude Spectrum (AS) based, Inter-Harmonic Phase Coherence (IHPC) and Inter-Cycle Phase Coherence (ICPC). The AS feature is close to the P2V measure proposed in \cite{HNM}. For a harmonic candidate in $\omega_p$, we consider the AS in the $[\omega_p-\frac{\omega_0}{2};\omega_p+\frac{\omega_0}{2}]$ span. A local Harmonic-to-Noise Ratio (HNR) is then estimated by calculating the subtraction (in dB) between the average level in $[\omega_p-\frac{\omega_0}{5};\omega_p+\frac{\omega_0}{5}]$, as it corresponds approximately to the main lobe, and the average level in the rest of the inital span.

The idea of IHPC is that the phase values across two consecutive harmonics should be similar, while this should not be the case for noisy contributions in speech. For this purpose, we use the Group Delay (GD) function, defined as $\frac{-d\phi(\omega)}{d\omega}$ where $\phi(\omega)$ is the unwrapped DFT phase spectrum. Note that a sufficiently high number of DFT points has to be used to facilitate phase unwrapping. The proposed IHPC is then expressed as the difference $GD(\omega_{p+1})-GD(\omega_{p})$ between two consecutive harmonic candidates.

The idea behind ICPC is to exploit the phase coherence for a particular harmonic when the windowing is shifted in time by the pitch period ($T_0$). Let us denote $\phi_1(\omega)$ the phase spectrum in the current frame, and $\phi_2(\omega)$ the phase spectrum after applying a delay of $T_0$ (and compensating the linear phase due to this delay). The differential phase is defined as $\Delta\phi(\omega)=\phi_1(\omega)-\phi_2(\omega)$. The ICPC feature which is used throughout this paper is expressed as the wrapped value of $\Delta\phi(\omega_{p+1})-\Delta\phi(\omega_{p})$.

\begin{figure}[!htpb]
  \centering
  \includegraphics[width=0.45\textwidth]{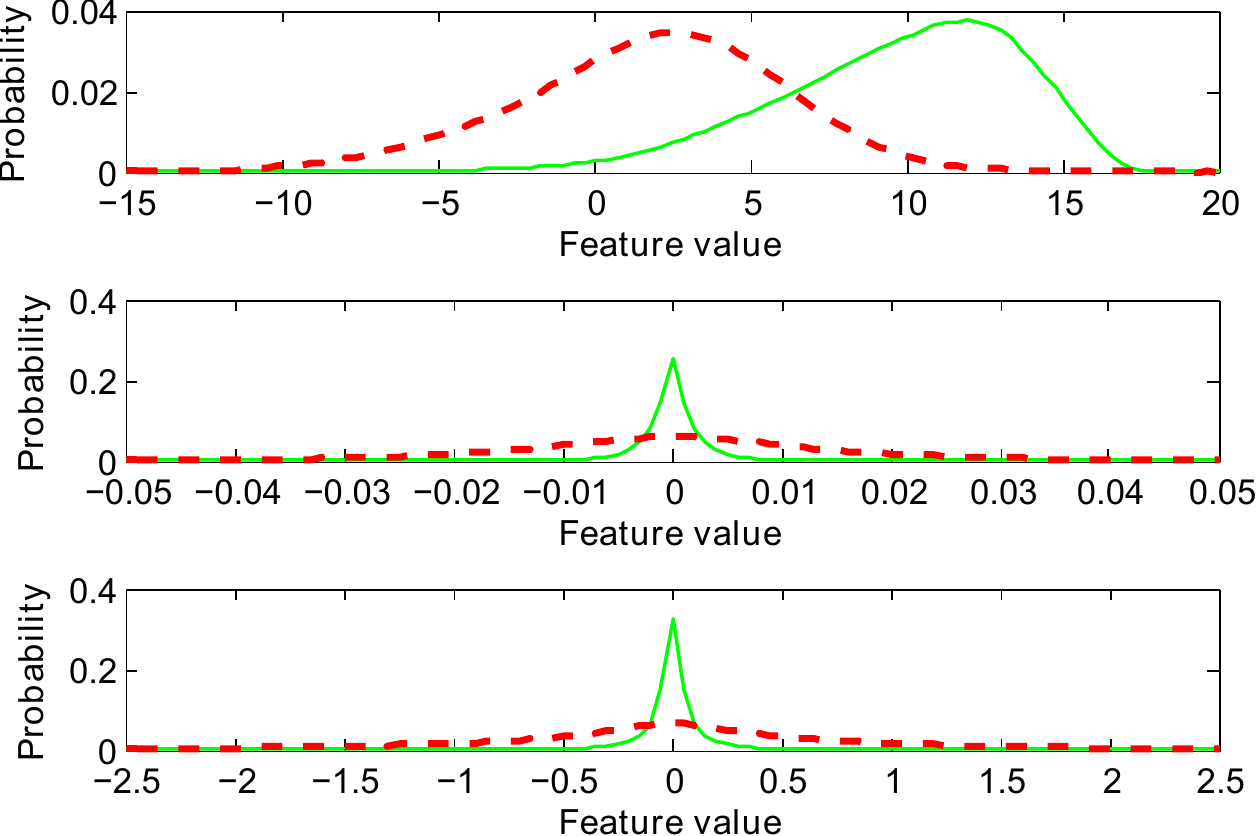}	
  \caption{Feature distributions for the 3 proposed features: AS (top plot), IHPC (middel plot) and ICPC (bottom plot). Histograms for real harmonics are in green bold line, while those of non-harmonics are in red dashed line.}
  \label{fig:FeatureHisto}  
	\vspace{-16pt}  
\end{figure}

Other formulations of the AS, IHPC and ICPC features are obviously possible. According to our experiments, however, the expressions proposed in this work led to a robust and discriminative characterization of the harmonicity at a given spectral peak. As an illustration, Figure \ref{fig:FeatureHisto} shows the distributions of these 3 features for harmonic candidates extracted on a development set of semi-synthetic signals (see Section \ref{ssec:Protocol}) for which the actual decision of harmonicity is known. It can be seen that the proposed features convey a relevant information to predict such a decision. It is also worth noting that the 3 proposed features are invariant to energy scaling.


\subsection{Decision Strategy}\label{ssec:Decision}
To find the most likely MVF, we consider the null hypothesis $H_0$ as a rejection of the harmonicity criterion. The estimated MVF is then:

\vspace{-6pt}  
\begin{equation}
MVF=\underset{\omega_m} {\mathrm{argmax}} [\prod_{k=1}^{m}{p(H_1|x_k,\omega_k}) \cdot \prod_{l=m+1}^{N_{harm}}{p(H_0|x_l,\omega_l})]
\label{eq:1}
\end{equation}

where $x_p$ and $N_{harm}$ respectively denote the measurement value and the number of harmonic candidates over the full band. The posterior for hypothesis $H_i$ can be written as:

\vspace{-6pt}  
\begin{equation}
p(H_i|x_p,\omega_p)=\frac{p(x_p|H_i,\omega_p)\cdot p(H_i|\omega_p)}{\sum_{j=0}^{1}{p(x_p|H_j,\omega_p)\cdot p(H_j|\omega_p)}}
\label{eq:2}
\end{equation}

Since the dependencies in $\omega_p$ are highly variable with the voice quality, speaker identity, recording conditions or voice type (e.g. speech vs singing voice), they will be omitted in the following. Considering both hypotheses to be equiprobable, it can be easily shown that (\ref{eq:1}) merely translates to the following ML criterion:

\vspace{-6pt} 
\begin{equation}
MVF=\underset{\omega_m} {\mathrm{argmax}} [\prod_{k=1}^{m}{p(x_k|H_1)} \cdot \prod_{l=m+1}^{N_{harm}}{p(x_l|H_0)}]
\label{eq:3} 
\end{equation}

In the rest of this paper, the strategy decision was made according to (\ref{eq:3}) where likelihoods were calculated by a Gaussian approximation of the feature distributions shown in Figure \ref{fig:FeatureHisto} (and which were obtained on the development set described in Section \ref{ssec:Protocol}). Note that in practice we worked with log-likelihoods for computational precision reasons.

\subsection{Time Smoothing}\label{ssec:Smoothing}
As for the P2V and SLM measures, the obtained MVF trajectories might contain spurious values and require some post-processing. For the objective evaluation (Section \ref{sec:Objective}, we applied a median filter of order 5 (considering a frame shift of 10 ms) in order to have results comparable with P2V. For the subjective evaluation, we preferred to use a moving average filter with a time constant of 30 ms (on both sides), since median filtering could lead to important jumps in the estimated contours, which could possibly lead to some auditory degradation. Finally, it is worth noting that, as it is the case for any other MVF estimation technique, the proposed method cannot distinguish between a production noise and a background noise. As a consequence, if speech is recorded in noisy conditions, MVF might be underestimated compared to what has been actually produced by the speaker.

\section{Objective Evaluation}
\label{sec:Objective}

This section aims at objectively comparing the proposed approach with P2V \cite{HNM} and SLM \cite{Erro}, for which we used the authors' original implementations. The experimental protocol and the results are further described in the following sections.


\subsection{Experimental Protocol}\label{ssec:Protocol}
The database used for the objective evaluation consists of \emph{semi-synthetic} signals obtained as follows. First, $F_0$ and 24 True-Envelope-based Cepstral Coefficients (TECC, \cite{TE}) were extracted from real recordings. $F_0$ was estimated using the Summation of the Residual Harmoncis (SRH) algorithm \cite{SRH}. TE is here required as standard cepstral analysis was reported in \cite{Singing} to be inappropriated for singing voice analysis. TE was estimated using the COVAREP toolkit \cite{COVAREP}. These files were further resynthesized using the DSM \cite{DSM} by imposing a constant value of the MVF. 7 versions of each file were created, with a fixed MVF ranging from 1 kHz to 7 kHz by step of 1 kHz. This way of doing allows us to have a ground truth to compare our MVF estimates with, and therefore to perform an objective assessment. Audio recordings were taken from CMU ARTCIC \cite{ARCTIC} speech databases (with 2 male and 2 female voices) and from the LYRICS singing voice database \cite{LYRICS} (with excerpts from 7 baritones, 3 counter-tenors and 3 sopranos). The whole database is made of 1400 files (across the 7 MVF values, and with a balance across speech and singing voice sounds) which were further split equally into a development and a test sets. The development set was used during the design of the proposed technique in Section \ref{sec:Proposed}, primarily to estimate the parameters of the Gaussian distributions used in the ML criterion. The test set was used to derive the results presented in Section \ref{ssec:Results}.

To assess performance, we use the rate of estimates with an error falling below a certain threshold $\theta$. By varying $\theta$ between 0 and an upper bound (here fixed to 1.5 kHz), an approximate Receiver Operating Characteristic (ROC) curve is obtained, which is usually summarized by a single metric: the Area Under the Curve (AUC). AUC values are then normalized such that an ideal method would yield an AUC of 1. This metric was preferred to the root mean squared error (RMSE), as RMSE would be excessively biased by gross estimation errors. 

\subsection{Results}\label{ssec:Results}
\begin{figure}[!htpb]
  \centering
  \includegraphics[width=0.45\textwidth]{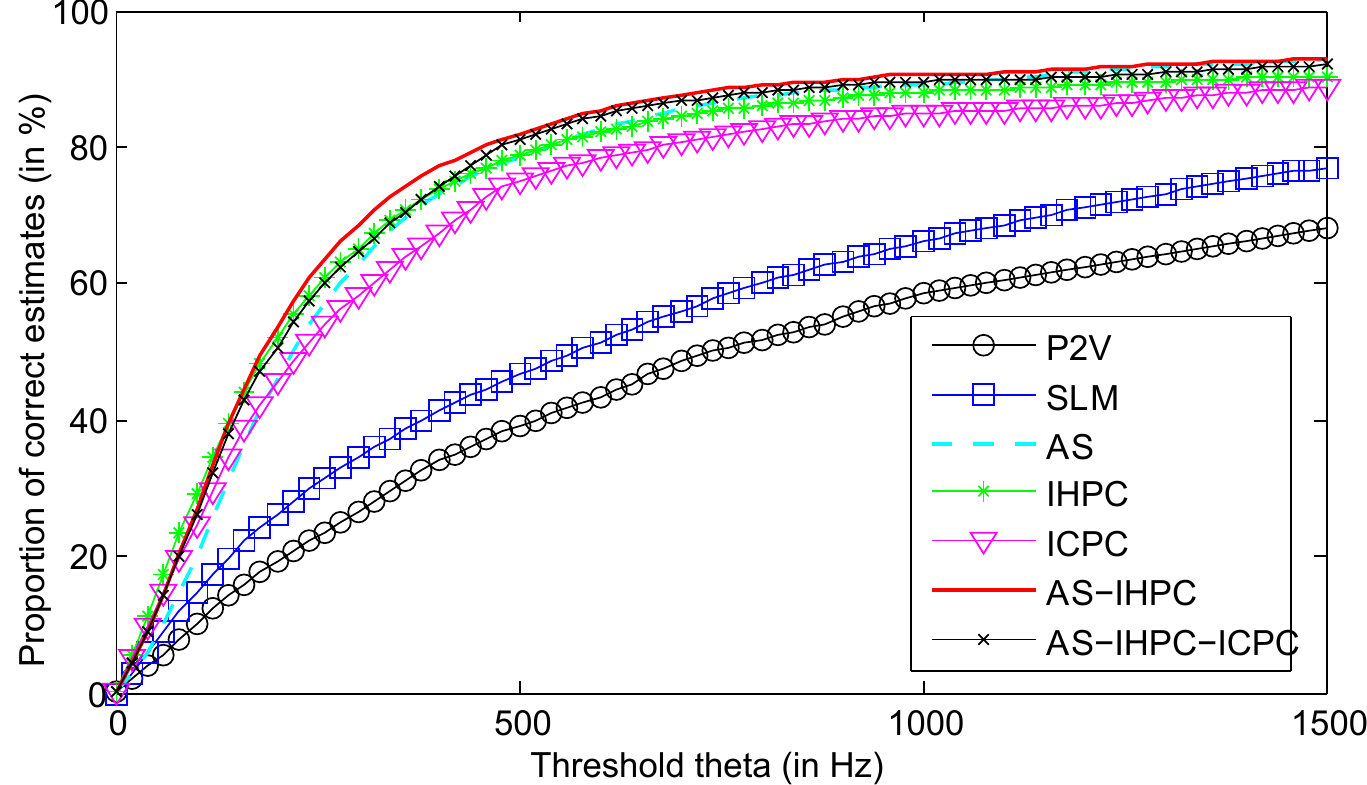}	
  \caption{Approximated ROC curve for the singing voice samples.}
  \label{fig:ROC_Singing}  
	\vspace{-8pt}  
\end{figure}

The approximate ROC curve obtained for the singing voice excerpts is shown in Figure \ref{fig:ROC_Singing}. The state-of-the-art techniques P2V and SLM are compared to 5 variants of the proposed method depending on whether the AS, IHPC and ICPC features are used alone or jointly. It can be first observed that the baseline techniques achieve a rather poor performance on singing voices, and that a great improvement is yielded with the proposed methods. Among these latter, the use of ICPC alone brings the lowest performance, while the best results are obtained by using AS and IHPC features jointly.

\begin{figure}[!htpb]
  \centering
  \includegraphics[width=0.48\textwidth]{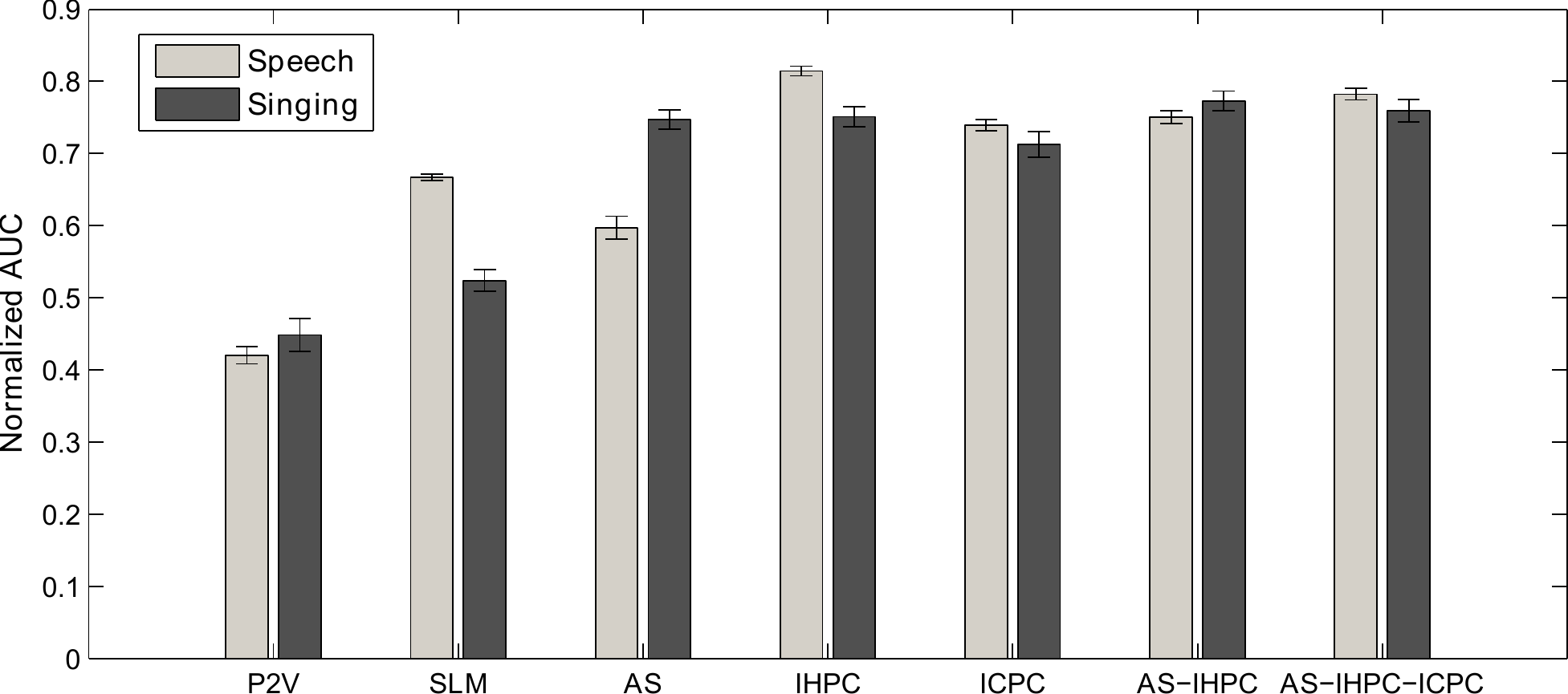}	
  \caption{Averaged AUC values for various MVF estimation techniques for speech and singing voice, together with their $95\%$ confidence intervals.}
  \label{fig:AUC_Results}  
	\vspace{-12pt}  
\end{figure}

Figure \ref{fig:AUC_Results} displays the AUC values averaged across all files for all compared methods. As it was observed in Figure \ref{fig:ROC_Singing}, the proposed techniques clearly outperform state-of-the-art methods in singing voice. Interestingly, it can be noted that phase-based features achieve a rather high performance in both speech and singing voice, contrary to the use of the AS feature for which a degradation is noticed in speech. We also observed informally the advantage of phase-based features over AS to be more pronounced in female than in male speech. Among the proposed features, it turns out that IHPC carries out the best results, followed by ICPC and AS. According to these experiments, the combination of IHPC with other features does not seem to bring much improvement. Nonetheless, as it will be discussed in Section \ref{ssec:Subj_Protocol}, we have observed on real signals that the joint use of several features tends to provide more stable and coherent estimates. 

\section{Subjective Evaluation}
\label{sec:Subjective}

Besides the objective evaluation on semi-synthetic signals, one might be interested in assessing the benefit of the proposed technique in the frame of voice synthesis. For this purpose we have carried out a subjective test whose details are given in Section \ref{ssec:Subj_Protocol} and whose results are described in Section \ref{ssec:Subj_Results}.

\subsection{Experimental Protocol}
\label{ssec:Subj_Protocol}
Again, our goal here is to cover a large diversity of voice signals. More precisely, six categories of voices are considered: male speech, female speech, child speech, baritones, counter-tenors, and sopranos. Adult speech samples are from AWB (male) and CLB (female) speakers of the CMU ARCTIC database \cite{ARCTIC}. Child speech sentences were from two 11 year-old female speakers, and were kindly provided by Acapela Group. In addition, samples of different singers were taken from the LYRICS database \cite{LYRICS}, for a  total of 13 singers (7 bass-baritones, 3 countertenors, and 3 sopranos). The whole database used in our subjective evaluation contains in total 66 recordings mixed across the 6 categories. Note that none of these files were part of the datasets used in Section \ref{sec:Objective}.

By visual inspection of the MVF values estimated by the proposed approaches on these real signals, we noticed that combining various features led, in some cases, to more stable estimates. We also observed that the AS-IHPC and AS-IHPC-ICPC estimates were extremely close. However, AS and IHPC features require the computation of a single FFT per frame, while ICPC requires one more FFT (for the next pitch period) and therefore roughly doubles the computational cost. For these reasons, AS-IHPC was chosen as a representative of the proposed approach in the following experiments.

As in Section \ref{ssec:Protocol}, sounds were resynthesized using DSM with 3 feature streams as input: $F_0$, 24 TECCs and the MVF estimated using either P2V, SLM or the proposed AS-IHPC technique. The evaluation consists of a Comparative Mean Opinion Score (CMOS) test, where participants are asked to compare two versions of the same file on a 7-point scale ranging from \textit{``much worse''} ($-3$) to \textit{``much better''} ($+3$). A nul score is given if both versions are found to be equivalent. In our experiments, two comparisons were considered: AS-IHPC vs. P2V, and AS-IHPC vs. SLM. Two groups of listeners took the test: 39 naive listeners (via the crowdsourcing website \emph{CrowdFlower}) and 13 speech experts. Each participant was presented 20 randomized pairwise comparisons.

\subsection{Results}
\label{ssec:Subj_Results}

\begin{figure}[!htpb]
  \centering
  \includegraphics[width=0.48\textwidth]{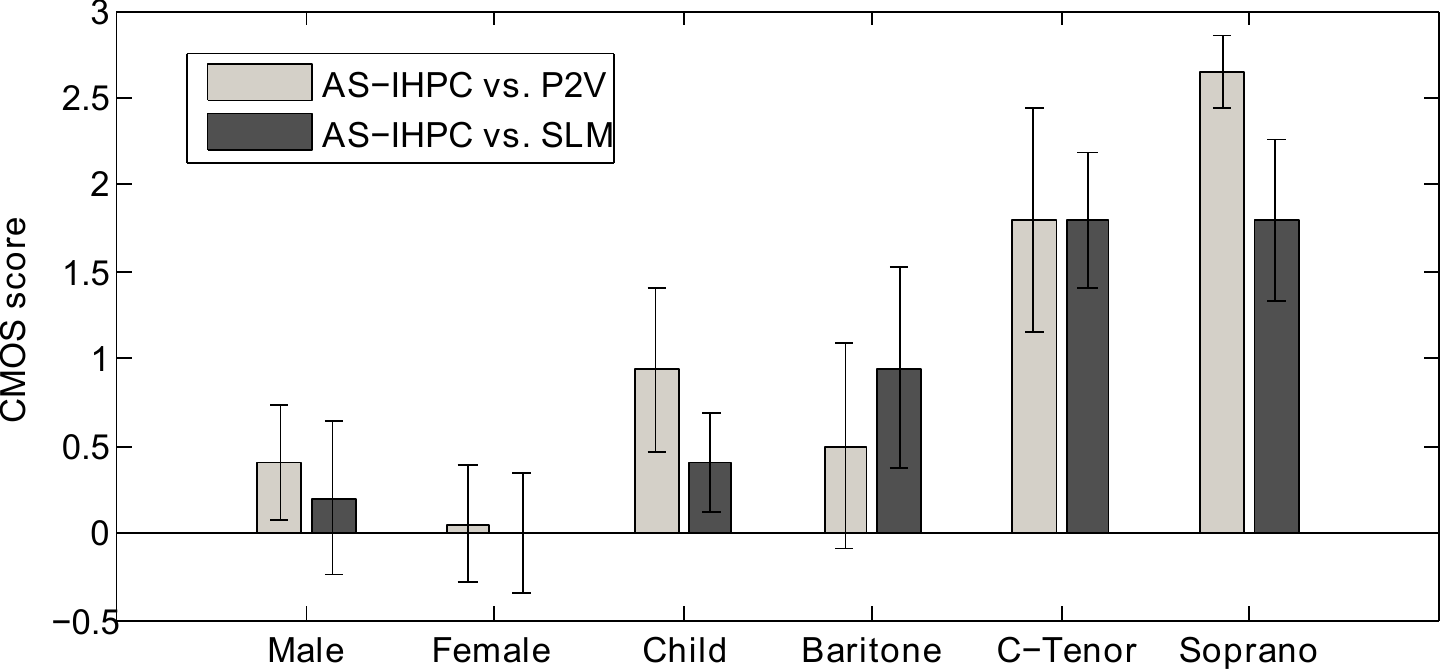}	
  \caption{Results of the CMOS test performed by the 13 speech experts. A positive score indicates a preference for the proposed AS-IHPC technique.}
  \label{fig:Subj_Experts}  
	\vspace{-6pt}  
\end{figure}

Results of the CMOS test performed by the speech experts are exhibited in Figure \ref{fig:Subj_Experts}. The general trend shows that the higher the pitch of the voice, the more the improvement of the proposed technique over the state of the art. Although the improvements in adult speech are in general not significant, AS-IHPC is clearly seen to outperform both P2V and SLM in child speech and singing voice samples. This improvement even becomes substantial for counter-tenors and sopranos, for which averaged CMOS scores above 1.8 are obtained (which roughly implies that AS-IHPC is found to be ``better'').

\begin{figure}[!htpb]
  \centering
  \includegraphics[width=0.48\textwidth]{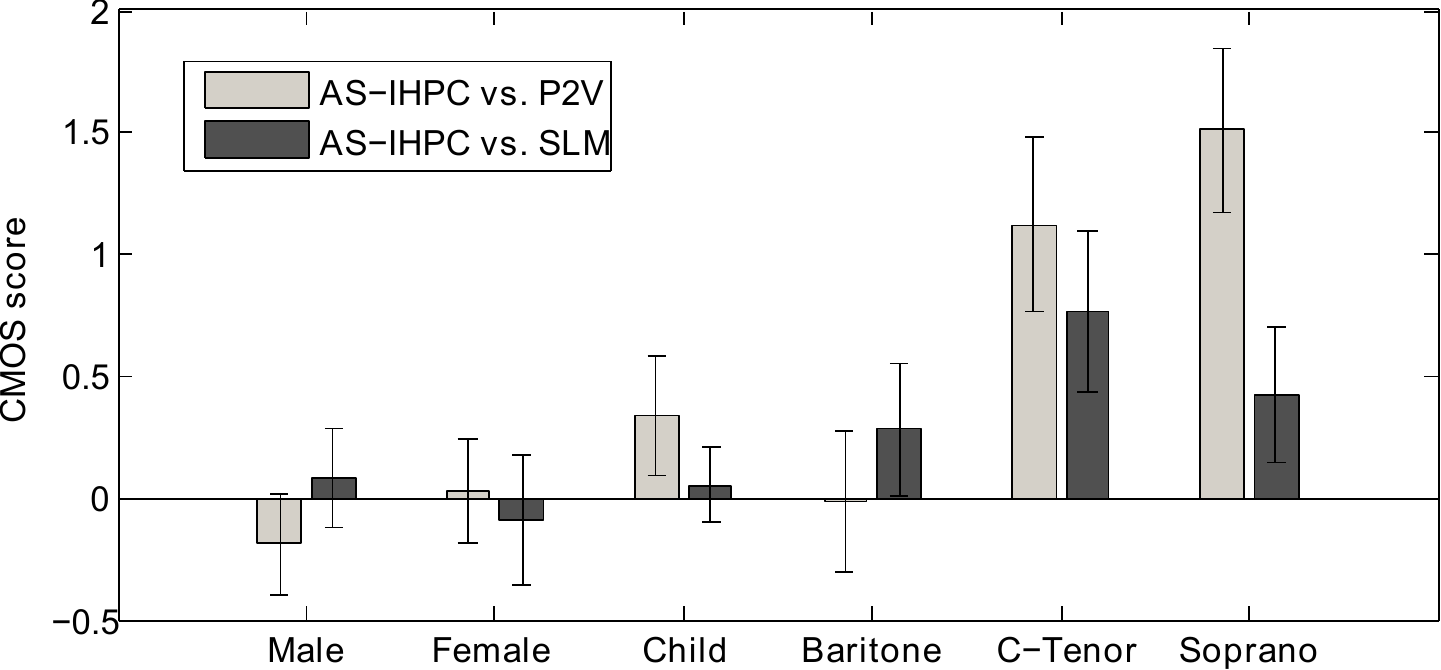}	
  \caption{Results of the CMOS test performed by the 39 naive listeners. A positive score indicates a preference for the proposed AS-IHPC technique.}
  \label{fig:Subj_Crowd}  
	\vspace{-12pt}  
\end{figure}

Results for naive listeners are presented in Fig. \ref{fig:Subj_Crowd}. Tendencies and conclusions drawn from these results are similar as with speech experts, except that they are in this case less pronounced. This can be explained by two facts: \emph{i)} compared to speech experts, naive listeners pay less attention to details; \emph{ii)} contrary to naive listeners, speech experts tend to use the full range of the CMOS scale. Nonetheless, it is appreciable to note that naive listeners also could observe differences between two versions of a voice sound which only differ by their estimated MVF contours.

\section{Conclusion}
\label{sec:Conclu}

This paper proposed a new approach for MVF estimation. This technique was shown to outperform two state-of-the-art methods both in an objective and subjective evaluation. In the perceptual assessment, a substantial improvement was observed for high-pitched voices (particularly for counter-tenors and sopranos). This was reported for both speech experts and naive participants. The success of the proposed method lies in its three novelties: \emph{i)} the use of the phase information, which is shown to have the potential to discriminate harmonicity; \emph{ii)} the joint use of features extracted from the amplitude and phase spectra; \emph{iii)} the use of a maximum likelihood criterion as a MVF decision strategy. A Matlab implementation of the proposed technique as well as audio examples can be found at \emph{tcts.fpms.ac.be/$\sim$drugman/Toolbox}.



\ifCLASSOPTIONcaptionsoff
  \newpage
\fi

\end{document}